# Student self-management, academic achievement: Exploring the mediating role of self-efficacy and the moderating influence of gender—insights from a survey conducted in 3 universities in America

Zhiqiang Zhao [1]*, Ping Ren [2], Qian Yang[3]

[1]*Beijing PhD Village Education Technology Co., Ltd; Beijing 100871, China*
[2] *Chengdu Ding Yi Education Consulting Co., Ltd, Chengdu 610023, China*
[3]*Beijing School, Tongzhou, Beijing, 101117, China*

* Corresponding: Zhiqiang Zhao (202128030258@mail.bnu.edu.cn)

**Abstract**
Excellent students are not only those who master more effective and efficient learning techniques to acquire and apply information. Even in the absence of correct learning, they are able to self-motivate, evaluate, and adjust their behavior. This study aims to explore the relationship between student self-management and academic achievement, with a focus on investigating the mediating role of self-efficacy and the moderating influence of gender in this relationship. A total of 289 students from three universities in the United States participated in this research. The results of the study indicate that students' level of self-management is positively correlated with their academic achievement, with self-efficacy playing a mediating role in this relationship and gender exerting a certain moderating effect. This study provides important insights into understanding the relationship between student self-management and academic achievement and supports the crucial role of educational leaders in educational practice.

**Keywords**：self-management, academic achievement, self-efficacy, educational leadership

**Introduction**
In the field of education, students' self-management skills have long been recognized as one of the key determinants of their academic achievement[1] [2]. Self-management encompasses students' abilities in goal setting, time management, self-monitoring, and other aspects of learning, crucial for helping them effectively organize learning tasks, overcome challenges, and maintain motivation [3]. At the same time, self-efficacy, as a critical concept in psychology, refers to individuals' confidence and beliefs in their ability to accomplish specific tasks[4] [5]. This belief not only influences individuals' performance when facing challenges but also affects their goal setting and level of effort. Therefore, self-efficacy plays a significant role in shaping students' beliefs about their own abilities and consequently influences their academic performance [6-9].

However, little attention has been paid to the mediating role of self-efficacy in this relationship and the moderating influence of gender. Building upon this, the present study aims to delve deeper into the relationship between student self-management and academic achievement [10]. Through analyzing survey data from three universities in the United States, several conclusions are revealed. Firstly, students' level of self-management is positively correlated with their academic achievement, indicating that students' abilities in effectively organizing their learning activities and managing their time are closely related to their academic performance[11].

Secondly, gender differences can influence the relationship patterns among student self-management, self-efficacy, and academic achievement [12-14]. Gender factors may lead to differences between male and female students in self-management and self-efficacy, thereby affecting their academic performance.

It is worth noting that in this relationship, educational leaders play a significant role. By providing support and resources, setting clear goals and expectations, and fostering a positive learning environment, they can promote the development of students' self-management and self-efficacy, thus enhancing their academic achievement [15, 16]. Understanding and effectively utilizing these relationships are of great significance for educational leaders. Through in-depth exploration in this study, we can better understand the relationship between student self-management and academic achievement and recognize the crucial role of educational leaders in educational practice.

**Self-management ability: The ability of students to self-regulate**

In broad terms, self-management encompasses a range of elements such as self-discipline, self-control, self-regulation, willpower, self-strength, and effort control[17]. It involves effectively regulating emotions, thoughts, satisfactions, and behaviors to inspire individuals to pursue their academic and personal goals[18, 19]. Furthermore, [20, 21] contend that self-management abilities not only aid individuals in understanding and predicting their emotions, thoughts, and behaviors but also assist in structuring and executing action plans to attain goals. Within the realm of education, self-management is defined as the capability to directly steer one's personality, behavior, and cognition towards accomplishing objectives or tasks. It serves as a vital tool across various educational settings, including academic courses and materials, and is indispensable for other realms and capacities. At its core, self-management offers effective strategies, procedures, and techniques to guide individuals' actions and behaviors successfully throughout the learning process. Through self-management, students acquire the skills to regulate emotions, set goals, and organize themselves, thereby becoming self-motivated individuals [22, 23]. By fostering resilient qualities through self-management, students can effectively confront challenges and difficulties encountered in their learning journey[11, 24]. Consequently, self-management not only influences individuals' competency levels but also positively impacts their perseverance in pursuing challenging objectives. Overall, self-management facilitates efficient learning, enabling students to adhere to their learning strategies while staying focused. Through the cultivation and enhancement of self-management abilities, individuals can better navigate the complexities of the learning process and achieve greater academic success.

**Self-efficacy: Students' own fueling station**

Self-efficacy is defined as an individual's belief in their ability to plan and execute the steps required to achieve specific goals [25]. The theory of self-efficacy also emphasizes that it refers to an individual's belief in their ability to plan and execute specific behavioral strategies to find solutions or complete tasks[26]. Students' self-efficacy refers to the belief in their ability to learn and perform behaviors at a particular level, influencing students' feelings, thoughts, and actions. Additionally, according to[27], students' high self-efficacy promotes the development of skills, the construction of abilities, and the cultivation of resilience by fostering task motivation and commitment, diligence, longer endurance, and adaptability. The importance of self-efficacy is also evident in its impact on academic achievement [28, 29]. The more confident individuals are in their abilities in a specific area, the higher their enthusiasm, involvement, and persistence when facing related learning tasks. This confidence not only makes individuals more willing to accept challenges but also enhances their expectations and anticipation of their learning outcomes[30, 31]. Therefore, a high level of self-efficacy is closely associated with individuals' academic success, promoting their continuous progress and achievements in academia. Furthermore, self-efficacy is not only a belief in one's own abilities but also an important factor influencing individuals' emotions, thoughts, and behaviors. It promotes the development of individual skills and abilities by enhancing task motivation and commitment while also positively impacting individuals' academic achievement. Thus, in education and training, cultivating and enhancing individuals' self-efficacy is crucial as it effectively promotes their overall development and enhances their academic achievement.

**Academic achievement: One of the most important elements students care about**
The concept of academic achievement holds broad meanings and significance in the field of education. It encompasses various grades, abilities, and skills students achieve within academic settings, as well as the impacts of these achievements on their learning and future development [32, 33]. Due to its complexity and breadth, the definition and understanding of academic achievement vary and are often categorized and explained from different perspectives. Firstly, academic achievement can be understood as students' performance and outcomes in structured academic environments. This includes their performance in classrooms, exam results, completion of assignments, and participation in extracurricular activities [34]. Additionally, academic achievement can be measured through standardized test results, research paper accomplishments, participation in research projects, and other criteria assessing students' academic abilities and achievements [35, 36]. Besides, academic achievement also reflects students' mastery of academic knowledge and skills, as well as their growth and development during the learning process [37, 38]. This involves not only understanding and mastering specific academic subjects but also the development of problem-solving abilities, critical thinking skills, creative thinking abilities, and other aspects. Therefore, academic achievement is not merely an evaluation of students' learning outcomes but also an assessment of their academic abilities and potential. Furthermore, academic achievement can also be measured through students' overall performance in school and social life. This includes the development of their leadership abilities, teamwork skills, social responsibility, and other aspects. Consequently, academic achievement reflects not only students' academic abilities but also their development and growth in comprehensive literacy [39, 40]. In conclusion, academic achievement is a comprehensive reflection of students' various achievements and performances in academic environments, covering their learning outcomes, academic abilities, comprehensive literacy, and other aspects. Understanding and evaluating this concept are crucial for the development of educational institutions and individual students, as it helps guide teaching practices and the formulation of students' learning strategies.

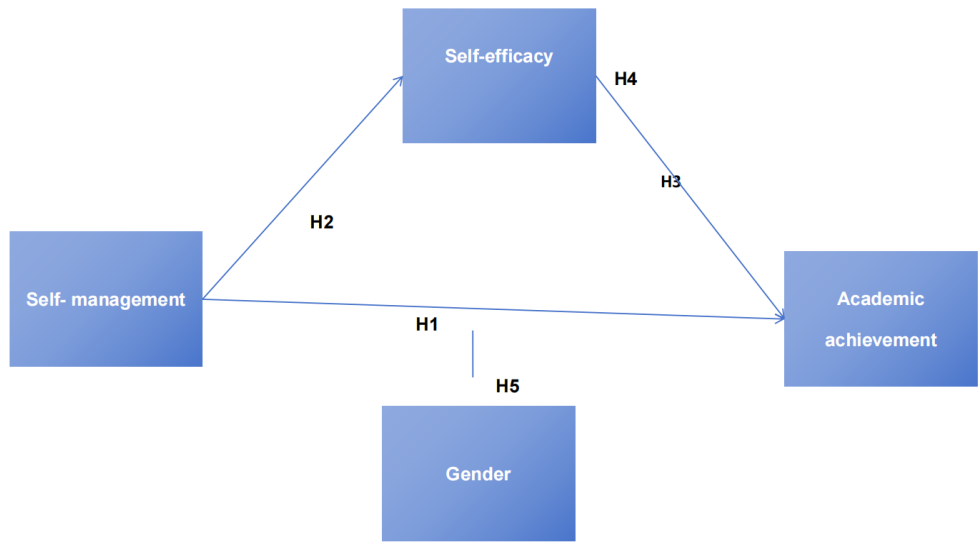

**Figure1. Variable framework of this research**

Figure 1 depicts the five hypotheses of this research:
Hypothesis 1: A positive correlation exists between students' self-management and academic achievement.
Hypothesis 2: A positive correlation exists between students' self-management and self-efficacy.
Hypothesis 3: A positive correlation exists between self-efficacy and academic achievement.
Hypothesis 4: Self-efficacy moderates the relationship between students' self-management and academic achievement.
Hypothesis 5: Gender moderates these relationships.

In summary, there is a close relationship between self-management, self-efficacy, and academic achievement, all of which are crucial for students' development. The core of this study lies in considering self-efficacy as a mediating variable and gender as a moderating variable, positioned within the relationship between self-management and academic achievement. To better understand these relationships, we have constructed a variable framework of this research, as depicted in Figure 1. The aim of this research design is to delve into the complex relationships among self-management, self-efficacy, gender, and academic achievement, providing valuable insights and guidance for educational practices and policy-making.

**Methods**
The independent variable of this study is students' self-management. The inspiration for this variable stem from the research by [41], which includes 11 items such as "I make a to-do list every day" and "I make an effort to complete tasks on time." We used a Likert five-point scale to assess these statements, with higher scores indicating stronger self-management abilities.

Additionally, we employed the "General Academic Self-Efficacy Scale" (GASE) developed by [42] to quantify students' self-efficacy. The survey utilized a five-point scale ranging from 1 to 5, representing strongly disagree to strongly agree. with higher scores indicating stronger self-efficacy feelings. When evaluating students' academic performance, we once again utilized the Likert five-point scale, with higher scores reflecting better academic performance. As for the moderating variable "gender," we classified it into two groups: male and female, based on traditional gender categorization. The online survey was administered in the form of a Google form. Among the 300 respondents received, we obtained 297 completed questionnaires, resulting in a response rate of 99%. Due to incomplete responses, 8 surveys were excluded, leaving a final questionnaire count of 289 for statistical analysis. The collection and processing of this data provide a reliable foundation for our in-depth analysis of the relationships between self-management, self-efficacy, gender, and academic achievement. Through meticulous surveys and data analysis, we aim to offer deeper insights and guidance for educational practices and policy-making.

## Results
### Summary of Sample Characteristics

**Table1. Summary of Sample Characteristics**

| Name | Item | Frequency | Percentage (%) | Cumulative percentage (%) |
|---|---|---|---|---|
| Gender | Female | 149 | 51.56 | 51.56 |
|  | Male | 140 | 48.44 | 100.00 |
| Age | 21-25 years old | 194 | 67.13 | 67.13 |
|  | 26-30 years old | 33 | 11.42 | 78.55 |
|  | More than 31 years old | 27 | 9.34 | 87.89 |
|  | Under 20 years of age | 35 | 12.11 | 100.00 |
| CGPA | 3.0-3.5 | 111 | 38.41 | 38.41 |
|  | 3.6-4.0 | 158 | 54.67 | 93.08 |
|  | Under 3.0 | 20 | 6.92 | 100.00 |
| Educational Level | Master | 67 | 23.18 | 23.18 |
|  | Phd student | 59 | 20.42 | 43.60 |
|  | Undergraduate | 163 | 56.40 | 100.00 |
| Major | Anesthesia | 1 | 0.35 | 0.35 |
|  | Animal science | 10 | 3.46 | 3.81 |
|  | Animation science | 4 | 1.38 | 5.19 |
|  | Art | 16 | 5.54 | 10.73 |
|  | Biology | 31 | 10.73 | 21.45 |
|  | Chemistry | 7 | 2.42 | 23.88 |
|  | Computer | 50 | 17.30 | 41.18 |

| Name | Item | Frequency | Percentage (%) | Cumulative percentage (%) |
|---|---|---|---|---|
| | Economics | 15 | 5.19 | 46.37 |
| | Economics policy | 19 | 6.57 | 52.94 |
| | Education | 7 | 2.42 | 55.36 |
| | Educational technology | 20 | 6.92 | 62.28 |
| | Forensic psychology | 1 | 0.35 | 62.63 |
| | Geology | 2 | 0.69 | 63.32 |
| | Higher education | 14 | 4.84 | 68.17 |
| | Lawyer | 12 | 4.15 | 72.32 |
| | Marxism | 2 | 0.69 | 73.01 |
| | Medicine | 9 | 3.11 | 76.12 |
| | Nurse | 2 | 0.69 | 76.82 |
| | Pe | 1 | 0.35 | 77.16 |
| | Physics | 27 | 9.34 | 86.51 |
| | Plant quarantine | 2 | 0.69 | 87.20 |
| | Precision surveying | 2 | 0.69 | 87.89 |
| | Preschool education | 5 | 1.73 | 89.62 |
| | Product design | 6 | 2.08 | 91.70 |
| | Psychology | 18 | 6.23 | 97.92 |
| | Satellite navigation | 2 | 0.69 | 98.62 |
| | Sociology of education | 4 | 1.38 | 100.00 |
| | Summary | 289 | 100.0 | 100.0 |

Based on the data presented in Table1, it is evident that the majority of students are at the undergraduate level, constituting 56.40% of the sample. The age distribution primarily centers around the 21-25 age range, accounting for 67.13% of the participants. These students represent diverse academic backgrounds, with 17.30% specializing in computer science, alongside others in fields such as education and business. Female students comprise 51.56% of the sample. Regarding Cumulative Grade Point Average (CGPA), a significant portion of the samples, 54.67%, fall within the "3.6-4.0" range, while 38.41% have a CGPA between 3.0 and 3.5, indicating generally strong academic performance across the sample.

**The relationship between students' self-management and academic achievement**

**Table2. Relationship results graph between self-management (iv) and academic achievement (dv)**

| | Self-management |
|---|---|
| Academic achievement | 0.987** |

|  | Self-management |
|---|---|

*p<0.05  ** p<0.01*

Based on the data in Table2, we can observe that the correlation coefficient between Self-management and Academic achievement is 0.987, exhibiting significant significance at the 0.01 level. This indicates a significant positive correlation between these two variables, suggesting that as self-management abilities improve, their academic achievement also increases correspondingly.

**The relationship between students' self-management and self-efficacy**

**Table3. Relationship results between self-management (iv) and self-efficacy (m1)**

|  | Self-management |
|---|---|
| Self-efficacy | 0.870** |

*p<0.05  ** p<0.01*

From the Table3, it can be observed that the correlation coefficient between Self-Management and Self-Efficacy is 0.870, and it demonstrates significance at the 0.01 level, indicating a significant positive correlation between these variables.

**The relationship between self-efficacy and academic achievement**

**Table4. Relationship results between self-efficacy (m1) and academic achievement (dv)**

|  | Self-efficacy |
|---|---|
| Academic achievement | 0.787** |

*p<0.05  ** p<0.01*

From Table4, it is evident that the strength of the relationship between the two variables is represented using Pearson correlation coefficient. The correlation coefficient value between the two variables is 0.787, showing significance at the 0.01 level, indicating a significant positive correlation between them.

**The mediating role of self-efficacy**

**Table 5. Mediation Effect Model Testing** *(n=289)*

|  | Self-efficacy | | | | | Academic achievement | | | | | Academic achievement | | | | |
|---|---|---|---|---|---|---|---|---|---|---|---|---|---|---|---|
|  | B | standard error | t | p | β | B | standard error | t | p | β | B | standard error | t | p | β |
| Constant | -0.000** | 0.000 | -11.339 | 0.000 | - | -1.934** | 0.043 | -44.731 | 0.000 | - | -1.948** | 0.051 | -38.310 | 0.000 | - |
| Self-management | 0.416** | 0.000 | 2699183348170378.563 | 0.000 | 1.000 | 0.098** | 0.001 | 103.027 | 0.000 | 0.987 | 0.068** | 0.000 | 6835937500000.000 | 0.000 | 0.689 |
| SELF |  |  |  |  |  |  |  |  |  |  | 0.068** | 0.000 | 6835937500000.000 | 0.000 | 0.287 |
| R 2 |  | 1.000 | | | | | 0.974 | | | | | 0.964 | | | |
| Adjust R 2 |  | 1.000 | | | | | 0.974 | | | | | 0.963 | | | |
| F value | F (1,287)=7.285590747055981e+28,p=0.000 | | | | | F (1,287)=10614.515,p=0.000 | | | | | F (2,286)=3795.134,p=0.000 | | | | |

* p<0.05  ** p<0.01

**Table 6. Indirect Effect Analysis Results**

| Item | Effect | Boot SE | BootLLCI | BootULCI | z | p |
|---|---|---|---|---|---|---|
| Self- management⇒Self-efficacy⇒Academic achievement | 0.028 | 0.048 | 0.369 | 0.558 | 0.596 | 0.551 |

*Note: BootLLCI refers to the lower limit of the Bootstrap resampled 95% interval, BootULCI refers to the upper limit of the Bootstrap resampled 95% interval. Bootstrap type: Percentile Bootstrap Method.*

Table5 provides a detailed display of three key mediation analysis models employed in this study. In subsequent research, researchers utilized Bootstrap resampling methods, conducting 5000 resamples to validate the mediation effects. The results presented in Table 6 are striking: for the "self-management⇒self-efficacy⇒academic achievement" mediation pathway, the 95% confidence interval range (95% CI: 0.369~0.558) does not include zero, indicating the existence and significance of this mediation pathway. These suggesting that the influence of self-management on academic achievement may be achieved through shaping individuals' self-efficacy.

**The moderating role of gender**

**Table 7. Results of Moderation Analysis with Gender as Moderating Variable *(n=289)***

|  | Model1 | | | | | Model2 | | | | | Model3 | | | | |
| --- | --- | --- | --- | --- | --- | --- | --- | --- | --- | --- | --- | --- | --- | --- | --- |
|  | *B* | standard error | *t* | *p* | *β* | *B* | standard error | *t* | *p* | *β* | *B* | standard error | *t* | *p* | *β* |
| Constant | 2.478 | 0.006 | 415.334 | 0.000** | - | 2.478 | 0.006 | 416.429 | 0.000** | - | 2.477 | 0.006 | 414.360 | 0.000** | - |
| Self-management | 0.098 | 0.001 | 103.027 | 0.000** | 0.987 | 0.098 | 0.001 | 103.063 | 0.000** | 0.988 | 0.098 | 0.001 | 96.982 | 0.000** | 0.991 |
| Gender |  |  |  |  |  | 0.019 | 0.012 | 1.586 | 0.114 | 0.015 | 0.019 | 0.012 | 1.613 | 0.108 | 0.015 |
| Self-management*Gender |  |  |  |  |  |  |  |  |  |  | -0.002 | 0.002 | -0.792 | 0.429 | -0.008 |
| *R 2* | 0.974 | | | | | 0.974 | | | | | 0.974 | | | | |
| *Adjust R 2* | 0.974 | | | | | 0.974 | | | | | 0.974 | | | | |
| *F value* | F (1,287)=10614.515,p=0.000 | | | | | F (2,286)=5336.552,p=0.000 | | | | | F (3,285)=3553.278,p=0.000 | | | | |
| *R 2* | 0.974 | | | | | 0.000 | | | | | 0.000 | | | | |
| *F value* | F (1,287)=10614.515,p=0.000 | | | | | F (1,286)=2.516,p=0.114 | | | | | F (1,285)=0.628,p=0.429 | | | | |

DV：Academic achievement
\* *p<0.05*  \*\* *p<0.01*

From Table 7, it can be seen that the moderation analysis consists of three models. Model 1 includes only the independent variable (Self-management). Model 2 adds the moderator variable (Gender) to Model 1, and Model 3 further includes the interaction term (the product of the independent variable and the moderator variable). For Model 1, the purpose is to investigate the impact of the independent variable (Self-management) on the dependent variable (Academic achievement) without considering the moderator variable (Gender). It can be observed from Table 7 that the independent variable (Self-management) shows significance (t=103.027, p=0.000<0.05). This implies that Self-management significantly influences Academic achievement.

Additionally, moderation effects can be examined in two ways. One way is to observe whether the change in the F-value from Model 2 to Model 3 is significant; the other way is to observe the significance of the interaction term in Model 3. In this analysis, the second method is utilized.

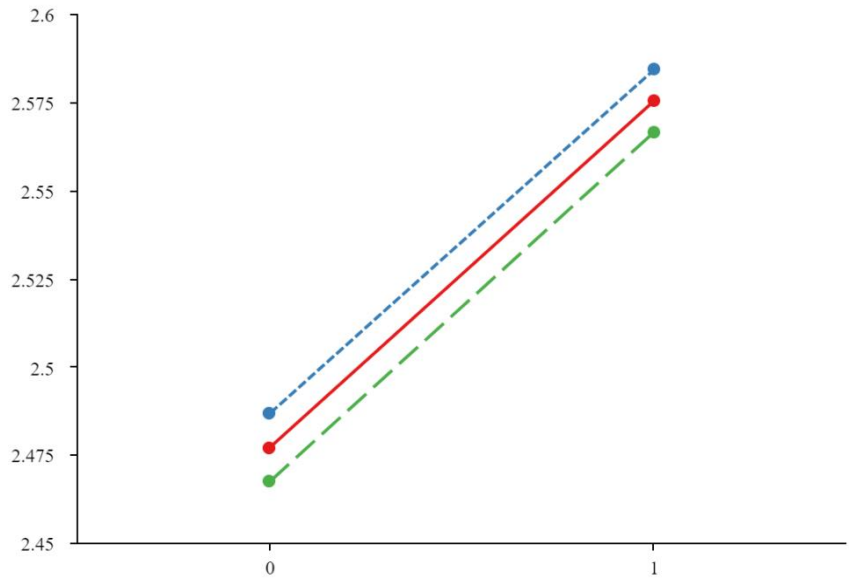

**Figure2.Moderation effects result**

From Table 7, it can be observed that the interaction term between Self-management and Gender is not significant (t=-0.792, p=0.429>0.05). Combined with the results of Model 1, which indicates a significant impact of Self-management on Academic achievement, it suggests that the effect of Self-management on Academic achievement remains consistent across different levels of the moderator variable (Gender), as shown in Figure 2.

**Discussion and conclusion**
In summary, the research reveals several key findings. Firstly, the results indicate a significant positive correlation between self-management and academic achievements. In other words, as students' self-management abilities strengthen, their academic achievements correspondingly improve. Secondly, the research demonstrates a positive correlation between students' self-management abilities and their self-efficacy. This suggests that enhancing students' self-management abilities can promote an increase in their self-efficacy. Furthermore, self-efficacy plays a positive role in promoting academic achievement, emphasizing the importance of psychological qualities in academic performance. It is noteworthy that the study finds self-efficacy to mediate the relationship between self-management abilities and academic achievement, implying that self-management abilities indirectly influence academic achievement by affecting self-efficacy. However, it is essential to emphasize that gender does not significantly moderate this association, indicating that the impact of self-management on academic achievement does not differ significantly across genders.

Combining the key findings, a series of measures can be taken to promote students' self-management skills and academic achievement. Firstly, according to[43-45], students should recognize the importance of self-management for academic success and actively cultivate self-management skills while enhancing self-efficacy. Parents play a crucial role in this process by providing necessary support and guidance to their children, encouraging them to adopt a positive mindset, and offering a conducive learning environment and resources. Similarly, teachers need to provide effective guidance and support to students, encouraging their participation in various learning activities and providing timely feedback and support. Lastly, educational administrators can offer training and support, create a favorable educational environment, and support relevant research and practices to enhance students' academic achievement and self-management skills[46, 47]. These comprehensive measures will help develop students' self-management abilities and improve their academic performance, thus establishing a conducive educational ecosystem and laying a solid foundation for their holistic development.

**Acknowledgments**
First and foremost, we wish to convey our heartfelt appreciation to the managing editor and associate editors for their tireless dedication and invaluable contributions to this publication. Secondly, we extend our sincere gratitude to the editorial team for their unwavering support and assistance throughout the entire process. Lastly, we would like to express our profound gratitude to the reviewers for their meticulous feedback, which has significantly enhanced the quality of this article.

**Funding**
The authors state that there is no funding

**Data availability statement**
The data that support the findings of this study are available from the first author upon reasonable request

**Declaration of interest statement**
The authors report there are no competing interests to declare